\documentclass[reprint, 10pt, a4paper, aip]{revtex4-1}
\usepackage[utf8]{inputenc}
\usepackage{amsmath}
\usepackage{amsfonts}
\usepackage{amssymb}
\usepackage{mathrsfs}
\usepackage{subfigure}
\usepackage[hidelinks]{hyperref}
\usepackage{outlines}
\usepackage{tikz}
\usepackage{booktabs}
\usepackage{float}

\newcommand{\bv}{\mathbf{v}}

\newcommand{\bw}{\mathbf{w}}

\newcommand{\D}{\mathrm{d}}

\begin{document}
	
\title{Fusion yield of plasma with velocity-space anisotropy at constant energy}
\date{\today}
\author{E. J. Kolmes}
\email[Electronic mail: ]{ekolmes@princeton.edu}
\author{M. E. Mlodik}
\author{N. J. Fisch}
\affiliation{Department of Astrophysical Sciences, Princeton University, Princeton, New Jersey 08540, USA}

\begin{abstract}
Velocity-space anisotropy can significantly modify fusion reactivity. 
The nature and magnitude of this modification depends on the plasma temperature, as well as the details of how the anisotropy is introduced. 
For plasmas that are sufficiently cold compared to the peak of the fusion cross-section, anisotropic distributions tend to have higher yields than isotropic distributions with the same thermal energy. 
At higher temperatures, it is instead isotropic distributions that have the highest yields. 
However, the details of this behavior depend on exactly how the distribution differs from an isotropic Maxwellian. 
This paper describes the effects of anisotropy on fusion yield for the class of anisotropic distribution functions with the same energy distribution as a 3D isotropic Maxwellian, and compares those results with the yields from bi-Maxwellian distributions. 
In many cases, especially for plasmas somewhat below reactor-regime temperatures, the effects of anisotropy can be substantial. 
\end{abstract}
	
\maketitle

\section{Introduction}

Velocity-space anisotropy of the particle distribution is naturally present in a broad variety of plasma devices. For instance, in a system with spatially nonuniform or time-varying magnetic fields, the conservation of the first and second adiabatic invariants often results in plasmas that are hotter or cooler in the directions perpendicular and parallel to the magnetic field.  

This anisotropy is a central part of the confinement scheme for devices like magnetic mirrors. However, it is also ubiquitous in devices that do not rely on it; for instance, tokamaks and stellarators can have significant velocity-space anisotropy, including whole populations of particles whose parallel energies vanish at the high-field regions of their orbits.\cite{Rome1979, LinLiu1995, Cooper2006, Cooper2009} Anisotropy can also be produced as part of the compression process in implosion devices like $Z$-pinches and $\theta$-pinches.\cite{Ochs2018ii}

In addition, many heating techniques deposit heat anisotropically. For instance, electron and ion cyclotron resonance heating both heat in the direction perpendicular to the magnetic field, whereas neutral beam heating preferentially heats in the direction of the beam.\cite{Choe1995, Maximov2004, Yamaguchi2005, Qu2014} For this reason, strongly driven plasmas are often observed to have anisotropic temperatures. 

Thermonuclear fusion requires producing very high-temperature plasmas. 
When designing fusion devices, the focus has traditionally been on increasing the Lawson ``triple product" of density, temperature, and confinement time. 
However, there are regimes in which velocity-space anisotropy can significantly modify fusion yields, providing an additional pathway to better fusion performance. 
In some limits, this fact is well-known; for instance, it has long been recognized that the fusion reactivity of a directed beam interacting with a bulk plasma must be treated differently than the interaction of Maxwellian ion populations.\cite{Mikkelsen1989, Towner1992, Hay2015ignition} 

The role of anisotropy in determining fusion yields has received less attention in contexts where the bulk plasma population is anisotropic (for instance, due to adiabatic invariance or auxiliary heating), though a few authors have explored different aspects of this problem. Kiwamoto \textit{et al.} reported that neutron counts from the GAMMA10 tandem mirror experiment appeared to be consistent with a 2D (rather than 3D) Maxwellian particle distribution in velocity space.\cite{Kiwamoto1996} 
Kalra, Agrawal, and Pandimani\cite{Kalra1988} explored bi-Maxwellian fusion reactivities numerically; Nath, Majumdar, and Kalra\cite{Nath2013} did similarly while also including tri-Maxwellians and net drifts. Both papers noted the same trend in bi-Maxwellian fusion reactivities that we will show in Figure~\ref{fig:anisotropicYields}. 

As a model distribution with which to understand anisotropy, the bi-Maxwellian has significant advantages: it is simple and, as we will show, the fusion yields of interacting bi-Maxwellians can be parameterized in a particularly straightforward way. 
By ``bi-Mawellian," we mean that ions are distributed in velocity space with  one temperature $T_{||}$  in say the parallel direction, and a different temperature $T_\perp$ in the perpendicular direction, where parallel and perpendicular are generally considered to be with respect to a magnetic field, but need not be. 
The intuitions that can be gained from studying bi-Maxwellian fusion yields can certainly be useful, and for some kinds of anisotropy -- for instance, anisotropy generated by auxiliary heating -- the bi-Maxwellian distribution function may capture reasonably well the velocity space anisotropy. 

In any event, as we will discuss in Section~\ref{sec:biMaxwellianYields}, the modification of the fusion yield can be understood as a combination of two distinct mechanisms:
First, the difference between $T_\perp$ and $T_{||}$ changes the orientations with which pairs of particles encounter one another. This mechanism should affect all anisotropic distributions, bi-Maxwellian and otherwise. 
Second, the balance between $T_\perp$ and $T_{||}$ affects the distribution of particle energies, regardless of their orientation, even if the total energy in the system remains fixed. 
This effect is particular to the bi-Maxwellian distribution, and will not in general affect other distributions in the same way. 

This second mechanism makes bi-Maxwellian distributions a problematic model for any source of anisotropy that does not change the overall distribution of particle energies. In particular, the conservation of adiabatic invariants (at least in the simplest, collisionless case, and in the absence of electrostatic potentials) changes the balance of the parallel and perpendicular energies while leaving the total energy of any given particle fixed. 

This leads to a question: to what extent is the behavior of the bi-Maxwellian fusion yield general? 
Can intuitions gained from the bi-Maxwellian case be safely applied to other anisotropic distributions? 
The purpose of this paper is to address these questions, and to clarify the role of anisotropy in fusion yields. 
Section~\ref{sec:biMaxwellianYields} describes and parameterizes the fusion yield for interacting bi-Maxwellian distributions. 
Section~\ref{sec:rotatedDistribution} introduces a class of distributions for which the degree of anisotropy is decoupled from the distribution of particle energies, and analyzes the associated fusion yield. Section~\ref{sec:gExample} shows an illustrative example using the distribution of particles in a collisionless mirror trap. Section~\ref{sec:discussion} discusses the implications of these results. 

\section{Fusion Yields from Interacting Bi-Maxwellian Distributions} \label{sec:biMaxwellianYields}

The fusion reaction rate between ions of species $a$ and $b$ can be written as $Y \doteq n_a n_b \langle \sigma w \rangle$, which can be expressed as 
\begin{align}
Y &= \int \D^3 \bv_a \, \D^3 \bv_b \, \sigma(w) w f_a(\bv_a) f_b(\bv_b). \label{eqn:yield}
\end{align}
Here $f_a$ and $f_b$ are the distribution functions of species $a$ and $b$, $\sigma(w)$ is the fusion cross-section, $\bw \doteq \bv_b - \bv_a$, and $w \doteq |\bw|$. 
In the case where the reaction is between two members of the same species, set $a = b$ and divide by two in order to avoid overcounting. 

Let $\hat b$ denote the unit vector in the direction of the magnetic field, and let $v_{s||} \doteq \hat b \cdot \bv_s$ and $\bv_{s \perp} \doteq \bv_s - \hat b v_{s||}$. The bi-Maxwellian distribution without net flows, given by 
\begin{gather}
f_s(\bv_s) = \frac{n_s}{T_{s||}^{1/2} T_{s \perp}} \bigg( \frac{m_s}{2 \pi} \bigg)^{3/2} \exp \bigg[ - \frac{m_s v_{s||}^2}{2 T_{s||}} - \frac{m_s \bv_{s \perp}^2}{2 T_{s \perp}} \bigg], \label{eqn:bimaxwellian}
\end{gather}
is arguably the simplest anisotropic variant of the Maxwellian. The family of bi-Maxwellians with the same total thermal energy can be parameterized by $\delta_s$, where 
\begin{align}
T_{s||} &= (1 + \delta_s) T_s \\
T_{s \perp} &= \bigg( 1 - \frac{\delta_s}{2} \bigg) T_s . 
\end{align}
In order to avoid negative temperatures, the $\delta_s$ parameter is assumed to fall between $-1$ and $2$. The distribution is isotropic when $\delta_s = 0$. The particle velocities are confined to the perpendicular plane when $\delta_s = -1$, and to the parallel axis when $\delta_s = 2$. For a pair of species $a$ and $b$, it is helpful to introduce a reduced mass 
\begin{gather}
\mu \doteq \frac{m_a m_b}{m_a + m_b}
\end{gather}
and inverse-mass-weighted temperatures 
\begin{align}
T_{||} &\doteq \frac{m_b T_{a ||} + m_a T_{b||}}{m_a + m_b} \\
T_\perp &\doteq \frac{m_b T_{a \perp} + m_b T_{a \perp}}{m_a + m_b} \, .
\end{align}
$T_{||}$ and $T_\perp$ can be parameterized in the same way as $T_{s||}$ and $T_{s \perp}$ using a weighted anisotropy parameter. If 
\begin{align}
T &\doteq \frac{m_b T_a + m_a T_b}{m_a + m_b} \label{eqn:T} \\
\delta &\doteq \frac{m_b T_a \delta_a + m_a T_b \delta_b}{m_b T_a + m_a T_b} \, ,
\end{align}
then 
\begin{align}
T_{||} &= (1 + \delta) T \\
T_\perp &= \bigg( 1 - \frac{\delta}{2} \bigg) T . 
\end{align}
In other words, the relative importance of the anisotropy of species $s$ is weighted by $T_s / m_s$. 

Plugging two bi-Maxwellian distributions into Eq.~(\ref{eqn:yield}) results in 
\begin{align}
Y &= \frac{n_a n_b}{T_{||}^{1/2} T_\perp} \bigg( \frac{\mu}{2 \pi} \bigg)^{3/2} \nonumber \\
&\hspace{30 pt} \times \int \D^3 \bw \, \sigma(w) w \exp \bigg[ - \frac{\mu w_{||}^2}{2 T_{||}} - \frac{\mu \bw_\perp^2}{2 T_\perp} \bigg] .
\end{align}
In terms of energy coordinates $\varepsilon_{||} \doteq \mu w_{||}^2/2$ and $\varepsilon_\perp \doteq \mu w_\perp^2/2$, with $\varepsilon \doteq \varepsilon_{||} + \varepsilon_\perp$, this is 
\begin{align}
&Y = n_a n_b \sqrt{ \frac{2}{\pi \mu T_{||} T_\perp^2} } \nonumber \\
&\hspace{0 pt} \times \int_{0}^{\infty} \D \varepsilon_{||} \int_0^{\infty} \D \varepsilon_\perp  \, \sigma(\varepsilon) \sqrt{\frac{\varepsilon}{\varepsilon_{||}}} \exp \bigg[ - \frac{\varepsilon_{||}}{T_{||}} - \frac{\varepsilon_\perp}{T_\perp} \bigg]. \label{eqn:energyYieldIntegral}
\end{align}
For fixed densities and particle masses, Eq.~(\ref{eqn:energyYieldIntegral}) specifies the yield $Y$ as a function of $T_{||}$ and $T_\perp$ -- or, equivalently, as a function of the temperature $T$ and the anisotropy parameter $\delta$. 
The expression can be simplified to get an integral over a single variable: 
\begin{align}
&Y = n_a n_b \frac{\sqrt{2}}{\sqrt{\mu T_\perp (T_\perp - T_{||})}} \nonumber \\
&\times \int_0^\infty \D \varepsilon \, \sigma(\varepsilon) \sqrt{\varepsilon} \exp \bigg[- \frac{\varepsilon}{T_\perp} \bigg] \text{erf} \bigg[ \sqrt{ \varepsilon \frac{T_\perp - T_{||}}{T_{||} T_\perp} } \bigg] . \label{eqn:simplifiedEnergyYieldIntegral}
\end{align}
Formally, Eq.~(\ref{eqn:simplifiedEnergyYieldIntegral}) is not defined for $\delta = -1$, $0$, or $2$, although it is well-behaved in these limits. When $\delta \rightarrow -1$, $T_{||}/T_\perp \rightarrow 0$ and 
\begin{align}
\lim_{\delta \rightarrow -1} Y 
&= \frac{2 n_a n_b}{3 T} \sqrt{\frac{2}{\mu}} \int_0^\infty \D \varepsilon \, \sigma(\varepsilon) \sqrt{\varepsilon} \exp \bigg[ - \frac{2 \varepsilon}{3 T} \bigg]. \label{eqn:biMaxwellianPerpYield}
\end{align}
In the opposite limit, where $\delta \rightarrow 2$ and $T_\perp / T_{||} \rightarrow 0$, 
\begin{align}
\lim_{\delta \rightarrow 2} Y &= n_a n_b \sqrt{ \frac{2}{3 \pi \mu T} } \int_0^\infty \D \varepsilon \, \sigma(\varepsilon) \exp \bigg[ - \frac{\varepsilon}{3 T} \bigg] . \label{eqn:biMaxwellianParYield}
\end{align}
Finally, in the isotropic limit, where $\delta \rightarrow 0$, let $Y_\text{iso} \doteq \lim_{\delta \rightarrow 0} Y$. Then $Y_\text{iso}$ is given by 
\begin{align}
Y_\text{iso} = \frac{2 n_a n_b}{T^{3/2}} \sqrt{\frac{2}{\pi \mu}} \int_0^\infty \D \varepsilon \, \sigma(\varepsilon) \varepsilon \exp \bigg[- \frac{\varepsilon}{T} \bigg] . \label{eqn:isoYield}
\end{align}

\begin{figure}
\includegraphics[width=\linewidth]{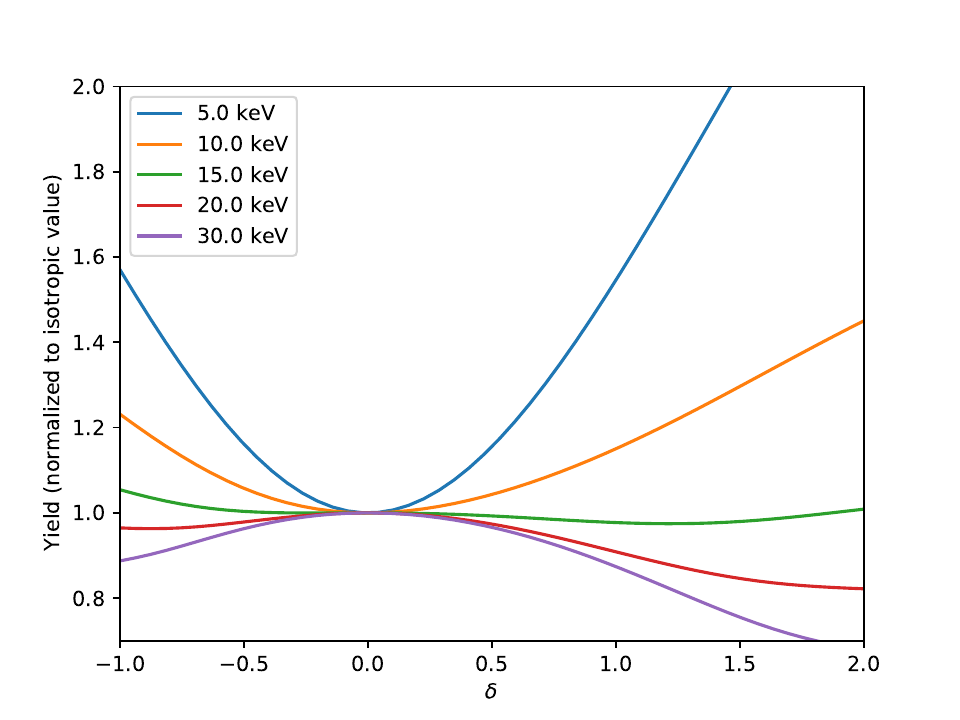}
\caption{The yield for a D-T plasma as a function of the anisotropy parameter $\delta$, evaluated for several choices of temperature. Below around 15 keV, less isotropic distributions have higher yields. Above about 15 keV, the reverse is true. }
\label{fig:anisotropicYields}
\end{figure}

Figure~\ref{fig:anisotropicYields} shows the dependence of $Y$ on $\delta$ for several choices of temperature $T$. 
The yields are normalized to the isotropic yield $Y_\text{iso}$ at the same temperature. 
The curves in the figure are for deuterium-tritium fusion; the cross-section $\sigma(\varepsilon)$ is modeled using the nine-parameter fit calculated by Bosch and Hale.\cite{Bosch1992} 
These numerical integrals were calculated using Gaussian quadrature; the higher-dimensional numerical integrals shown in the following section use a mix of quadrature and the VEGAS Monte Carlo algorithm.\cite{Lepage1978}

The dependence of the yield on anisotropy depends dramatically on the temperature. For colder temperatures, less isotropic distributions produce higher yields. As the temperature increases, the effect of anisotropy becomes less dramatic, until at around 15 keV it reverses sign. Then, as the temperature increases, isotropic temperatures produce the highest yields by increasingly large margins. This trend was previously pointed out by Kalra, Agrawal, and Pandimani\cite{Kalra1988} and Nath, Majumdar, and Kalra.\cite{Nath2013} 

\begin{figure}
\includegraphics[width=\linewidth]{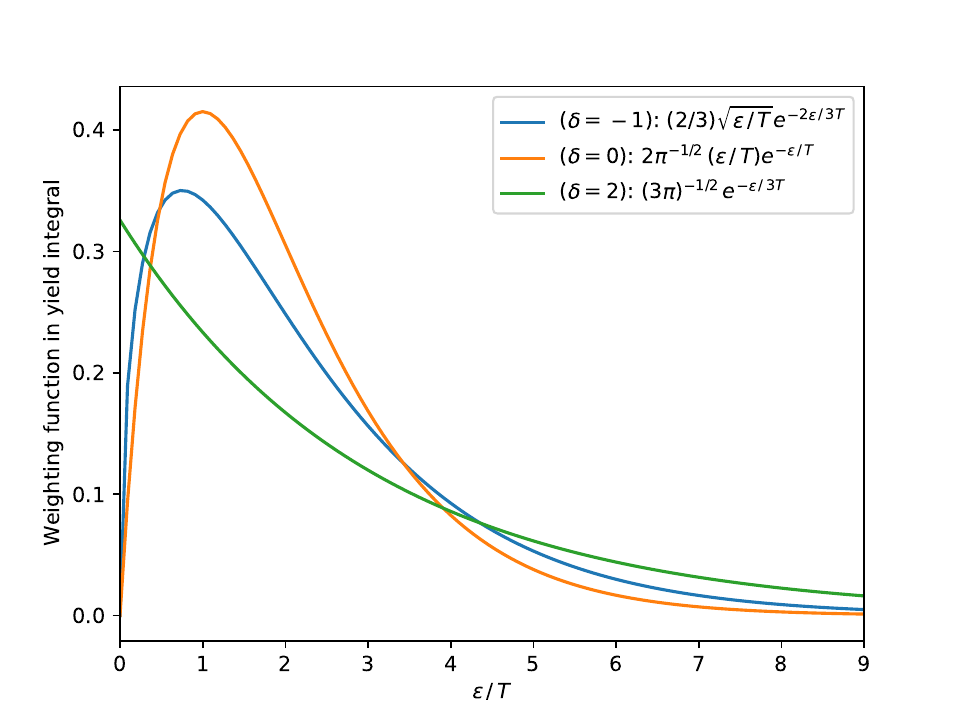}
\caption{The functions by which the cross-section is weighted in the yield integrals described by Eqs.~(\ref{eqn:biMaxwellianPerpYield}), (\ref{eqn:biMaxwellianParYield}), and (\ref{eqn:isoYield}). They can be interpreted as the product of $\sqrt{\varepsilon/T}$ and the distribution of the center-of-mass energies of pairs of particles. The less isotropic distributions have comparatively large numbers of particles in the highest-energy parts of the tails. }\label{fig:integrands}
\end{figure}

From one perspective, the explanation for this trend is relatively straightforward. The yield integral can be understood as the average of $\sigma(\varepsilon) \sqrt{2 \varepsilon / \mu}$ over the distribution of velocity differences between pairs of particles. The shape of that distribution depends on $\delta$, and can be seen (weighted by $\sqrt{\varepsilon}$ and a constant factor) in Figure~\ref{fig:integrands} for $\delta = -1$, $0$, and $2$. The less isotropic distributions have more particles at the highest-energy parts of the tails, whereas the more isotropic distribution has more particles in the moderately high-energy region. The highest-energy parts of the tails are most important at lower temperatures, since the cross-sections are steeper functions of energy at lower energies. 

However, this does not necessarily resolve the more basic question: why does the distribution of velocity differences depend on anisotropy in this way? There are two effects at play. 

To some extent, this behavior can be understood in terms of the alignment of the velocities of different particles. In a less isotropic distribution (larger $|\delta|$), the velocities are increasingly confined either to the perpendicular plane or to the parallel axis, depending on the sign of $\delta$. When the velocities are confined to a smaller-dimensional space, a given pair of velocities is increasingly likely to be aligned (or anti-aligned) rather than orthogonal. In other words, less isotropic distributions have a greater number of ``head-on" collisions, whereas more isotropic distributions have more ``side-swipe" collisions. 

As an example, consider two particles with the same mass and with speeds $v_1$ and $v_2$. If the two velocities are oriented along the same direction, then their relative velocity will be either $v_1 + v_2$ or $|v_2 - v_1|$, depending on whether they are aligned or anti-aligned. If the two velocities are completely orthogonal to one another, then their relative velocity will be $\sqrt{v_1^2 + v_2^2}$. 

For a relatively cold plasma, fusion reactions are very rare events, and they are much more likely to happen for those pairs of particles with the largest relative velocities. In terms of fusion yields, this incentivizes ``high-risk, high-reward" configurations in which velocities are confined to lower-dimensional subspaces; some pairs of particles will be moving in the same direction and have low relative velocities, but some will have very high-energy head-on collisions. 

On the other hand, for a hotter plasma, fusion events are less rare. Pairs of particles with orthogonal velocities may not reach the relative velocities that are possible in a head-on collision, but they will reliably have relative velocities higher than the velocity of either individual particle. At higher temperatures, where the fusion cross-section is a less steep function of energy, these orthogonal orientations are more favorable. 

However, this cannot be the whole story. Consider a fusion reaction between species $a$ and $b$ for which $T_a / m_a \gg T_b / m_b$. In this case, the orientation of the particles does not matter; species $b$ is effectively immobile, and all of the relative motion is provided by the thermal motion of species $a$. But the yield integrals depend on the species' anisotropies only through the dimensionless parameter $\delta$, which can take on a full range of values between $-1$ and $2$ even if one species is immobilized. In other words, all of the effects of anisotropy can apparently be observed even in a regime in which the relative orientation of the particles does not matter whatsoever. 

To see how this is possible, consider the dependence of $Y$ on $f_a$ and $f_b$, as shown in Eq.~(\ref{eqn:yield}). The anisotropy of the distribution functions affects $Y$ in two ways. First, for any given $v_a = |\bv_a|$ and $v_b = |\bv_b|$, it changes the likelihood of a given $w$ by modifying the relative orientation of $\bv_a$ and $\bv_b$. This leads to the effect described above. But $\delta$ can also affect the distribution of single-particle energies (i.e., the distributions of $v_a$ and $v_b$ themselves). Bi-Maxwellian distributions with different values of $\delta$ have differently shaped tails in energy space. This is much like the effect shown in Figure~\ref{fig:integrands}, but for single-particle energy distributions rather than the distribution of energies of pairs of particles. 

Again, though, this is not sufficient on its own to explain all of the effects of anisotropy. For instance, note that $\delta$ depends on both $\delta_a$ and $\delta_b$, and that the resulting combined $\delta$ parameter is quite different depending on whether or not $\delta_a$ and $\delta_b$ have the same sign. It is even possible for two anisotropic single-particle distributions to have $\delta = 0$. This would not happen if it was only the dependence of the single-particle energy distributions on $\delta$ that mattered. 
It appears, then, that the observed behavior of $Y(\delta, T)$ must follow from a combination of two things: that anisotropy changes the relative orientation of pairs of particle velocities, and that it changes the shapes of the single-particle energy distributions. 

Distinguishing between the two mechanisms is important because there are physical effects that create anisotropy while leaving the energy distribution unchanged. Consider a system in which the first adiabatic invariant $m v_\perp^2 / 2 B$ is conserved. If a particle moves through a region with varying field strength $B$, in the absence of collisions or potentials, the total energy of the particle remains fixed, and it trades $v_{||}^2$ with $v_\perp^2$ in order to conserve the adiabatic invariant as $B$ varies. 
It is not obvious to what extent the fusion yield in this system will behave like the bi-Maxwellian yield, because it is not obvious how to distinguish between the two mechanisms affecting the bi-Maxwellian. 

\section{A Different Maxwellian Analog} \label{sec:rotatedDistribution}

\begin{figure*}[t]
	\centering
	\includegraphics[width=\linewidth]{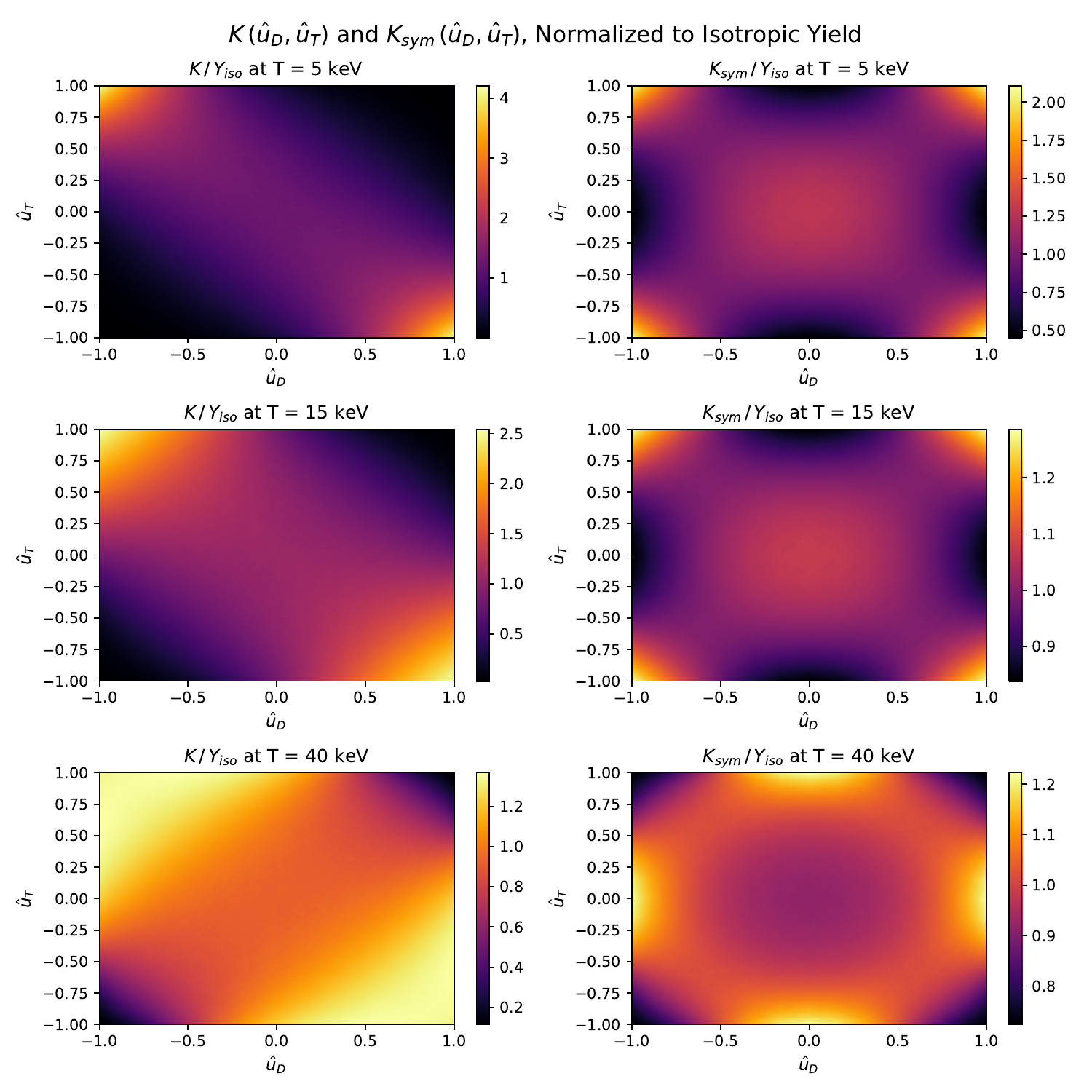}
	\caption{This figure shows $K$ and $K_\text{sym}$ as they compare to the isotropic Maxwellian yield $Y_\text{iso}$ for a variety of temperatures. The cases shown here use the deuterium-tritium cross-section and assume that the deuterium and tritium are at the same temperature.}\label{fig:combinedKernel}
\end{figure*}

In order to disentangle these two effects, it is helpful to construct a class of distribution functions which exhibit one mechanism but not the other. To that end, consider distributions of the form 
\begin{gather}
f_s(\bv_s) = g_s(\theta_s, \phi_s) h_s^M(v_s), \label{eqn:gDistribution}
\end{gather}
where $(v_s, \theta_s, \phi_s)$ are a spherical coordinate system:
\begin{gather}
\bv_s = \hat x \, \sin \theta_s \cos \phi_s + \hat y \, \sin \theta_s \sin \phi_s + \hat z \, \cos \theta_s ,
\end{gather}
$h_s^M(v_s)$ is a Maxwellian velocity distribution: 
\begin{align}
h_s^M(v_s) \doteq n_s \bigg( \frac{m_s}{2 \pi T_s} \bigg)^{3/2} e^{-m_s v_s^2 / 2 T_s} \, ,
\end{align}
and $g_s(\theta_s, \phi_s)$ is normalized such that 
\begin{gather}
\int_0^{2 \pi} \D \phi_s \int_0^\pi \D \theta_s \, \sin \theta_s \, g_s(\theta_s, \phi_s) = 4 \pi. 
\end{gather}
When $g_s = 1$, Eq.~(\ref{eqn:gDistribution}) gives a 3D isotropic Maxwellian. For any other choice of $g_s$, the distribution will have that same single-particle energy distribution, but different choices of $g_s$ can still introduce anisotropy and can therefore change the statistics of how pairs of particles' velocities align. 
In addition to being a useful example, this kind of distribution can act as a simple model for anisotropy due to the conservation of an adiabatic invariant, since varying $g$ modifies the distribution of pitch angles without changing the kinetic energy of any particle. 

In many cases of interest, a distribution function may be anisotropic due to dependence on $\theta_s$ but not $\phi_s$, since in a magnetized plasma any dependence on $\phi_s$ is averaged out over a Larmor gyration. Let $u_s \doteq \cos \theta_s$, so that $\D^3 \bv_s = v_s^2 \, \D v_s \, \D \phi_s \, \D u_s$. 
Define $K(\hat u_a, \hat u_b)$ by 
\begin{align}
&K(\hat u_a, \hat u_b) \doteq 4 \int \D^3 \bv_a \D^3 \bv_b  \nonumber \\
&\hspace{30 pt} \sigma(w) w h_a^M(v_a) h_b^M(v_b) \delta( u_a - \hat{u}_a ) \delta( u_b - \hat{u}_b ) . \label{eqn:K}
\end{align}
The factor of 4 appears because the integral of $g_s(u_s)$ over $u_s$ is normalized to 2. Then for any $g_a(u_a)$ and $g_b(u_b)$, 
\begin{align}
Y = \frac{1}{4} \int \D \hat u_a \, \D \hat u_b \, K(\hat{u}_a,\hat{u}_b) g_a(\hat u_a) g_b(\hat u_b) . 
\end{align}
Understanding $K(\hat{u}_a, \hat{u}_b)$ makes it possible to understand the range of behaviors that can be attained for more general distributions of the form $g_s(u_s) h_s^M(v_s)$. 

$K(0,0)$ can be understood as the yield in which each species has a 3D Maxwellian distribution with all velocity vectors rotated to lie on the $\hat x - \hat y$ plane. This distribution is analogous to the bi-Maxwellian with $\delta = -1$. In both cases the velocity vectors lie entirely on the $\hat x$-$\hat y$ plane, but the former has the $v$ distribution of a 3D Maxwellian whereas the latter (the $\delta = -1$ bi-Maxwellian) has the structure of a 2D Maxwellian. 

The analog to the $\delta = 2$ bi-Maxwellian is $K_\text{sym}(1,1)$, where the symmetrized $K$ is defined by 
\begin{align}
K_\text{sym}(\hat u_a, \hat u_b) \doteq \frac{1}{2} &\big[ K(\hat u_a, \hat u_b) + K (\hat u_a, - \hat u_b) \big]  . 
\end{align}
$K(\hat{u}_a, \hat{u}_b)$ is already symmetric with respect to the exchange of $\hat{u}_a$ and $\hat{u}_b$, so this symmetrized function is sufficient to describe the yields for distributions that satisfy $g_s(u_s) = g_s(-u_s)$. 
$K_\text{sym}(1,1)$ is the yield due to 3D Maxwellian distributions with all of their vectors rotated to lie on the $\hat z$ axis (with equal numbers in either direction). 
In some ways, it makes the most sense to compare $K_\text{sym}$ with the bi-Maxwellian yields (rather than $K$), since the imposed symmetry ensures that the distribution functions do not have net axial particle flows. 
Of course, $K$ and $K_\text{sym}$ are identical when $\hat{u}_a = \hat{u}_b = 0$. 

\begin{figure*}
	\centering
	\includegraphics[width=\linewidth]{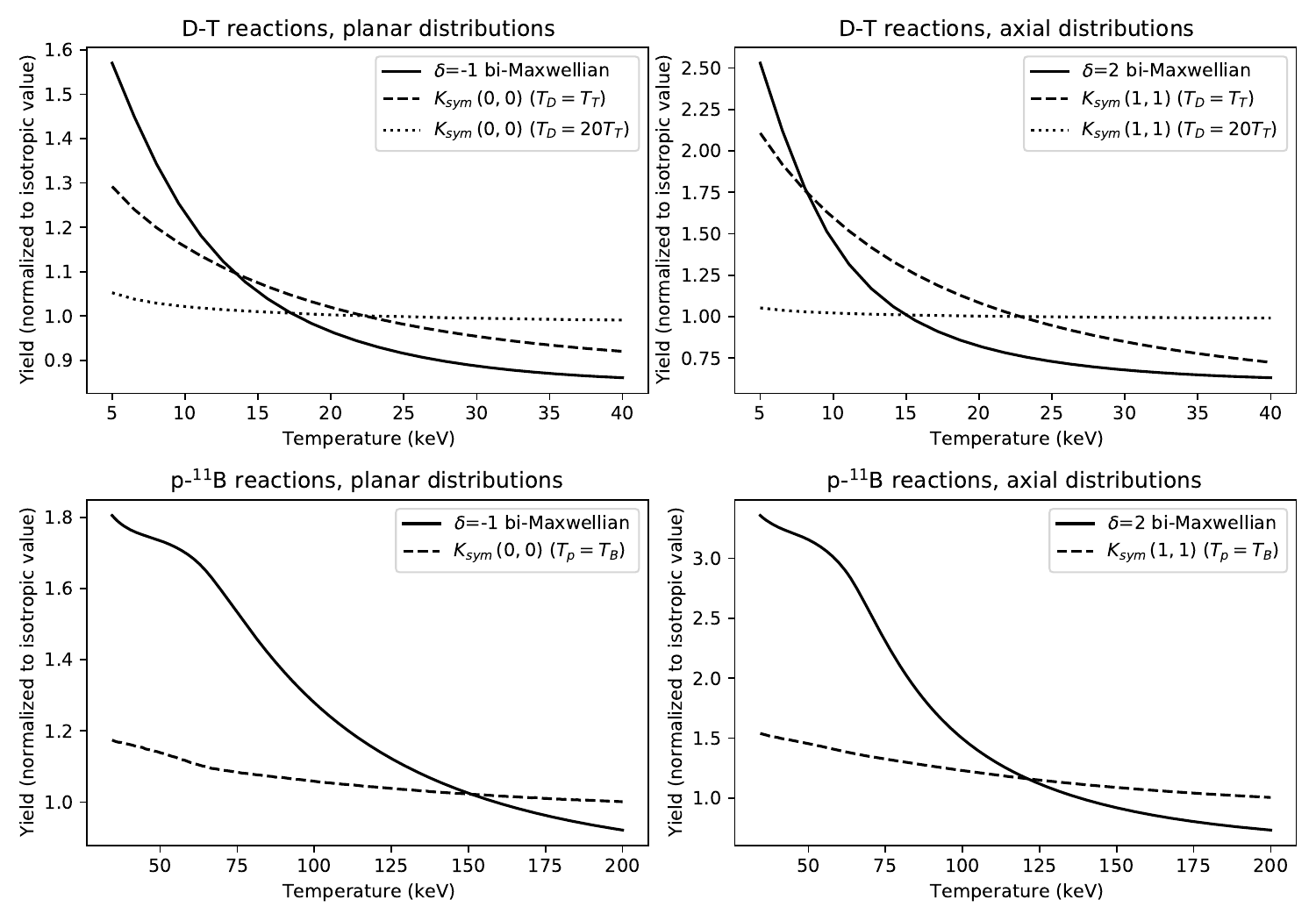}
	\caption{These plots show the fusion yield in a variety of scenarios as a function of temperature (normalized to the yield of isotropic Maxwellian distributions at the same temperature). The upper panels are for D-T reactions and the lower panels are for p-$^{11}$B. The dashed lines indicate $K_\text{sym}$ in which the two species have equal temperatures; the dotted lines in the D-T plots indicate the case in which $T_D = 20 T_T$. In cases where the species' temperatures are not equal, the $x$-axis shows the inverse-mass-weighted average of the two. For the unequal-temperature scenario shown here, that means $T_D = (50/31) T$ and $T_T = (5 / 62) T$.} \label{fig:combinedComparison}
\end{figure*}

Figure~\ref{fig:combinedKernel} shows $K$ and $K_\text{sym}$ as functions of $\hat{u}_D$ and $\hat{u}_T$ for deuterium-tritium fusion at several different temperatures. At low temperatures, $K_\text{sym}$ is largest when $\hat{u}_D$ and $\hat{u}_T$ are near $\pm 1$ or $0$. At higher temperatures, the trend reverses and $K_\text{sym}$ is smallest at these choices of $\hat{u}_D$ and $\hat{u}_T$. This follows from the same orientation argument made in Section~\ref{sec:biMaxwellianYields}; these distributions are confined to smaller-dimensional subspaces in which pairs of particle velocities are comparatively likely to be aligned or anti-aligned rather than orthogonal, and these configurations are more favorable for increasing the fusion yield at lower temperatures. 

If it is not symmetrized, $K$ does not show this same behavior, because it distinguishes between configurations in which pairs of particles will be aligned and those in which they will be anti-aligned. This is determined by the relative signs of $\hat{u}_D$ and $\hat{u}_T$. If it is possible to guarantee that pairs of particles will have oppositely oriented velocities, then the highest yields will be attained by the counter-propagating 1D velocity distributions. 

Figure~\ref{fig:combinedComparison} shows $K_\text{sym}(0,0)$, $K_\text{sym}(1,1)$, the $\delta = -1$ bi-Maxwellian yield, and the $\delta = 2$ bi-Maxwellian yield for a range of temperatures. The yields in the figure are normalized to the yield $Y_\text{iso}$ of an isotropic 3D Maxwellian at the same temperature. In all cases, the anisotropic distribution outperforms the isotropic distribution at lower temperatures and underperforms at higher temperatures. This effect is most pronounced for the 1D distributions. For the highest and lowest temperatures, it is more pronounced for the bi-Maxwellian yields than for $K_\text{sym}(0,0)$ or $K_\text{sym}(1,1)$. 
However, there is a range of intermediate temperatures for which $K_\text{sym}(0,0)$ and $K_\text{sym}(1,1)$ are further away from $Y_\text{iso}$ than are their bi-Maxwellian counterparts. 

There are limits in which $K_\text{sym}$ differs more dramatically from its bi-Maxwellian counterparts -- for instance, when the thermal velocities of the reacting species are more disparate. 
Recall that the relative thermal velocities of species $a$ and $b$ do not affect the bi-Maxwellian yields, so long as the combined $T$ parameter given by Eq.~(\ref{eqn:T}) remains fixed. 
This is not the case for $K$ or $K_\text{sym}$. 
Note, for example, that in the limit where $T_a / m_a \gg T_b / m_b$, $K$ will no longer depend on the choices of $g_a$ or $g_b$. This can be seen in Eq.~(\ref{eqn:K}), since $w \rightarrow v_a$ in this limit. 

Perhaps the most practical example in which this distinction would become important is when the reacting species have disparate masses -- for instance, in p-$^{11}$B fusion. 
This is shown in the lower two panels of Figure~\ref{fig:combinedComparison}; as expected, there is a larger disparity between the two kinds of anisotropy for this case. 
The numerical calculations shown in the figure use the piecewise cross-section fit described by Nevins and Swain.\cite{Nevins2000}
p-$^{11}$B provides a convenient formal example because of the large mass disparity between the reactants, but the possibility of enhancing the fusion reactivity also happens to be particularly topical. Ignition in p-$^{11}$B fuel is difficult for a plasma in which all species have equal-temperature isotropic Maxwellian distributions, and different ways in which the fusion yield might exceed that of the isotropic Maxwellian case have received significant interest in the recent literature.\cite{Lalousis2012, Hora2015, Eliezer2016, Putvinski2019, Hora2020} 

However, comparing the effects of anisotropy on two entirely different reactions is somewhat messy, since the cross-sections for different reactions can vary widely. 
An alternative way to demonstrate the same physics is to calculate the yields for D-T reactions in which the two species have very different thermal velocities. This can be accomplished either by having $T_D \neq T_T$ or by artificially modifying their masses (in fact, for the purposes of yield calculations, the two are formally equivalent). Consider a scenario in which 
\begin{align}
T_D &= \bigg( \frac{m_D + m_T}{\alpha m_D + m_T} \bigg) T \\
T_T &= \alpha T_D 
\end{align}
and the masses are left at their natural values. 
The inverse-mass-weighted $T$ given by Eq.~(\ref{eqn:T}) is unchanged, so the bi-Maxwellian yields are the same for any choice of $\alpha$. The dashed lines in the upper panels of Figure~\ref{fig:combinedComparison} show $K_\text{sym}$ when $\alpha = 1$ and the dotted lines show $K_\text{sym}$ when $\alpha = 1/20$. When $\alpha$ is far from 1, the effects of anisotropy on $K_\text{sym}$ are strongly suppressed, and $K_\text{sym}(\hat{u}_a, \hat{u}_b) \rightarrow Y_\text{iso}$ for all choices of $\hat{u}_a$ and $\hat{u}_b$. 

\section{Example: Yield for a Collisionless Mirror} \label{sec:gExample}

\begin{figure*}
	\centering
	\includegraphics[width=.49\linewidth]{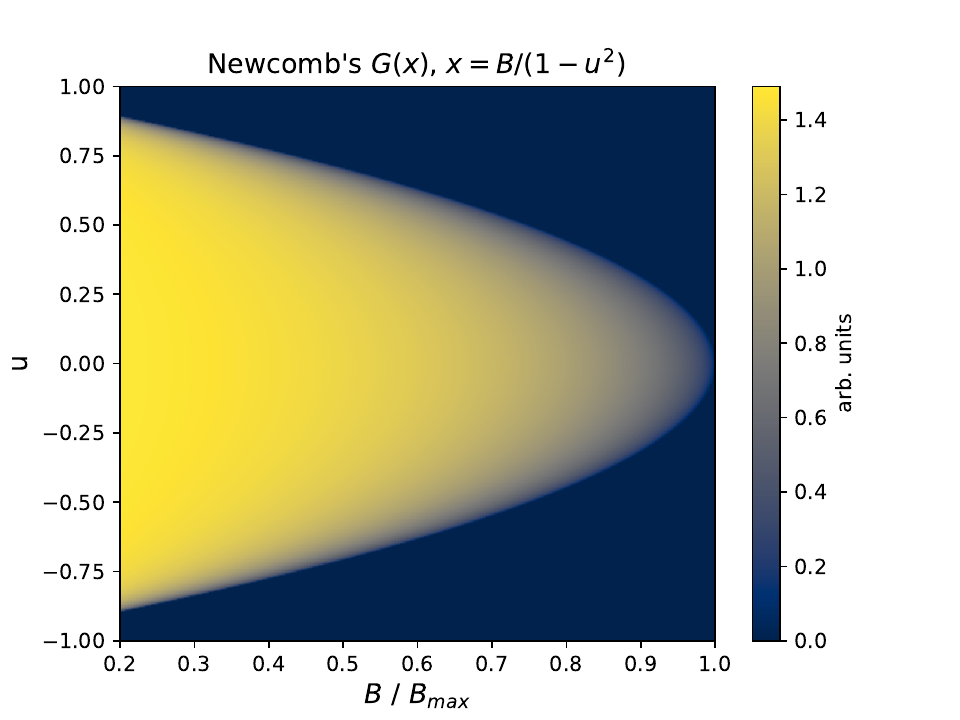}
	\includegraphics[width=.49\linewidth]{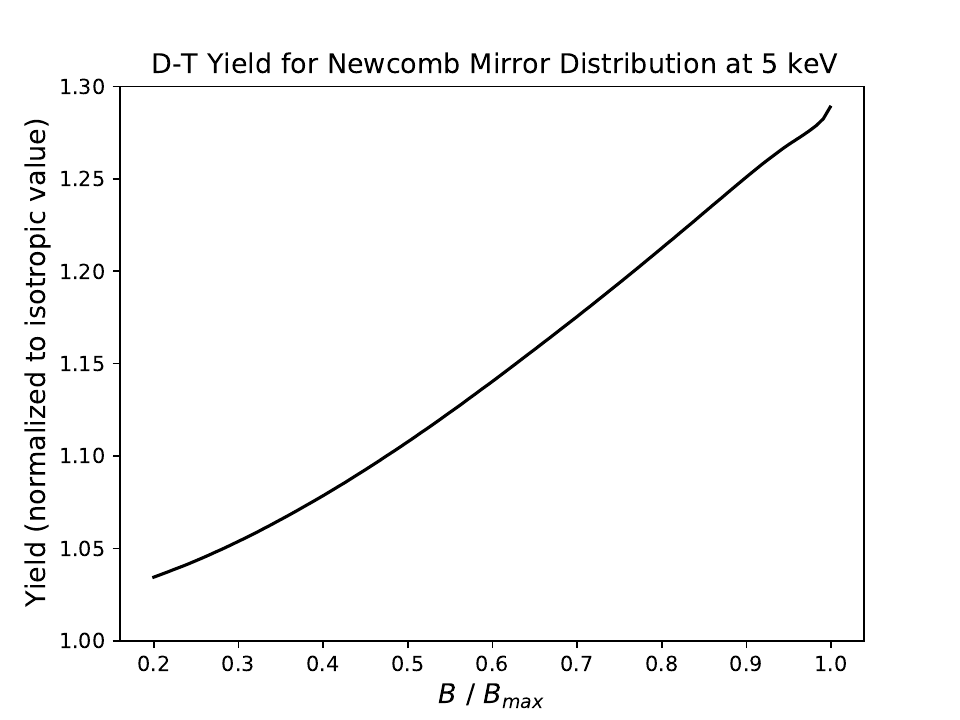}
	\caption{The left panel shows Newcomb's angular distribution $G$ as a function of $B / B_\text{max}$ and $u$. The right panel shows the yield for a D-T plasma at 5 keV with Newcomb's collisionless mirror distribution, normalized to the yield of an isotropic distribution at the same local temperature and density. } \label{fig:newcomb}
\end{figure*}

As can be seen in Figure~\ref{fig:combinedKernel}, the extremal values of $K_\text{sym}(\hat{u}_a, \hat{u}_b)$ tend to be fairly localized in $(\hat{u}_a, \hat{u}_b)$ space. The yield for a particular choice of $g_a(u_a)$ and $g_b(u_b)$ is determined by a weighted average of $K_\text{sym}(\hat{u}_a, \hat{u}_b)$ over $\hat{u}_a$ and $\hat{u}_b$; it is natural to wonder how much the yield would still be modified after this averaging in a real system. This could be done for any $g_a(u_a)$ and $g_b(u_b)$, but it may be helpful to see an example. 

With that in mind, consider the following equilibrium angular distribution for a long, thin mirror, obtained by Newcomb in the collisionless limit.\cite{Newcomb1981} Denote the maximum magnetic field strength by $B_\text{max}$. Define 
\begin{align}
G(x) &\doteq \frac{3 B_\text{max} + 2 x}{\pi} \sqrt{ x (B_\text{max} - x) } \nonumber \\
&\hspace{30 pt}+ \frac{3 B_\text{max}^2}{\pi} \arctan \sqrt{ \frac{B_\text{max}}{x} -1 } \label{eqn:newcombG}
\end{align}
for $0 \leq x \leq B_\text{max}$, and set $G(x) = 0$ otherwise, as per Newcomb's Eq.~(C16). 
Then, for the notation and normalization used in Section~\ref{sec:rotatedDistribution}, the angular distribution $g(u)$ is given by 
\begin{gather}
g(u) = 2 G \bigg( \frac{B}{1-u^2} \bigg)  \bigg[ \int_{-1}^{+1} \D u \, G \bigg( \frac{B}{1-u^2} \bigg) \bigg]^{-1} . \label{eqn:newcombg}
\end{gather}
The local field strength $B$ can be understood as a parameterization of axial position. 
Newcomb's solution is separable -- that is, the full kinetic distribution can be written as the product of $g(u)$ and a function of $v$ -- and was derived for a case without any electrostatic potential. For present purposes, it provides a good example of an equilibrium with the intuitively expected behavior for a collisionless mirror trap. The angular distribution is plotted in the left panel of Figure~\ref{fig:newcomb}. Near the ends of the trap, $B$ approaches $B_\text{max}$, and an increasing proportion of particles' velocities are oriented in the direction perpendicular to the field. 

For this example distribution, the effect of anisotropy on D-T fusion yields is shown in the right panel of Figure~\ref{fig:newcomb}. The angular distribution in the figure is given by Eq.~(\ref{eqn:newcombg}), and the distribution of speeds is given by a Maxwellian at 5 keV. Note that the yields in the figure are normalized to the yield of an isotropic Maxwellian at the same local density and temperature, so the plot of the variation of the yield with $B$ shows the effects of anisotropy but not the effects of axial variation in density. 

The yield results in Figure~\ref{fig:newcomb} show a substantial yield improvement (about 30\%) over an isotropic plasma near the end-coils of the mirror, but a much reduced enhancement in the middle of the mirror. These results can be compared with the plot of $K_\text{sym}$ at 5 keV in Figure~\ref{fig:combinedKernel}. When $B$ is close to $B_\text{max}$, the yield is comparable to $K_\text{sym}(0,0)$ (a purely perpendicular distribution), as one would expect. For regions with smaller $B$, the distribution is more isotropic and the modification of the yield decreases accordingly. The dropoff in the yield enhancement is approximately linear in $B$. Depending on the axial profile of $B$ for a particular mirror trap, this suggests that it might still be possible to observe enhancement in the fusion yields some distance away from the end-coils. Note that this particular example distribution's anisotropy tends to result in excess energy in the perpendicular rather than parallel direction. As can be seen in Figure~\ref{fig:combinedKernel}, a larger enhancement could be possible at the same temperature if the distribution instead favored excess parallel energy. 

\section{Discussion} \label{sec:discussion}

The different types of anisotropic distribution considered in this paper differ in significant ways. 
One of the major conclusions of this paper is that one must be careful when comparing anisotropy from different sources, and that subtle differences between different distributions can have surprisingly large impacts on the behavior of the fusion yield. 
For instance, in the limit in which one fuel ion has a much larger thermal velocity than the other, anisotropy can substantially modify the fusion yield from a pair of bi-Maxwellian distributions, but it no longer has any effect on the $g_s h_s^M$ distributions discussed in Section~\ref{sec:rotatedDistribution}. 
However, it is equally important to recognize that there are some ways in which the dependence of the fusion yield on anisotropy appears to be quite general, if not universal. 
This can be understood in terms of the relative likelihoods that pairs of particles' velocities will be aligned, anti-aligned, or orthogonal. 

The yields of $g_s h_s^M$ distributions are of particular interest for systems that are anisotropic due to the conservation of adiabatic invariants (this includes mirror machines, but trapped particle effects can also be important in toroidal devices). 
These devices will not necessarily have distributions with exactly this form; electrostatic and centrifugal potentials can prevent the kinetic energy of a particle from being constant over the course of an orbit, as can collisions and particle losses.\cite{Lehnert1971, Pastukhov1974, Chernin1978, Cohen1978, Bekhtenev1980, Zhang2019} 
Nonetheless, these distributions capture an essential characteristic of this kind of anisotropy: that the mechanism of anisotropy generation itself does not change the distribution of particle energies. 

These results have applications for optimizing and predicting fusion yields in plasma devices. 
Anisotropy can affect the fusion yield most dramatically in cases in which the plasma is hot enough to be producing fusion events but cooler than the threshold for ignition. 
One practical takeaway of this paper is that anisotropy (of either of the two types discussed) can strongly enhance fusion reactivities for relatively low-temperature devices. 
To the extent that anisotropy can be engineered -- for instance, by modifying the magnetic field coil configuration -- this could allow for greater fusion yields without the need to increase the plasma temperature or density. 

These effects become smaller and eventually reverse sign at higher temperatures. 
For $\delta = -1$ and $\delta = 2$ bi-Maxwellians, the D-T fusion yield becomes equal to $Y_\text{iso}$ at $T \approx 14 \text{ keV}$ and 15~keV, respectively. This reversal is pushed to higher temperatures for the analogous values of $K_\text{sym}(1,1)$ and $K_\text{sym}(0,0)$: $T \approx 23 \text{ keV}$ and 22 keV, respectively. 
Of course, for other fusion reactions these crossing points are different; for the p-$^{11}$B cases shown in Figure~\ref{fig:combinedComparison}, the reversal is not until the plasma gets to temperatures well over 100~keV. 

Especially for the kind of anisotropy generated by the conservation of adiabatic invariants (i.e., modeled by an integral over $K_\text{sym}$), these effects could still be nontrivial at reactor-relevant temperatures. For a deuterium-tritium plasma at 15 keV, $K_\text{sym}(1,1) / Y_\text{iso} \approx 1.29$ and $K_\text{sym}(0,0) / Y_\text{iso} \approx 1.07$. The yield enhancement for a real device would presumably be an average over some finite region of $(\hat{u}_D, \hat{u}_T)$-space, but these results suggest that significant enhancements are still possible at these temperatures. 
In a device as large and costly as a magnetic fusion reactor, even a relatively small enhancement to the reactivity is worth noting. 

This paper has considered two possible classes of particle distributions: separable distributions of the form $g_s(\theta_s,\phi_s) h_s^M(v_s)$ (for cases in which the mechanism generating anisotropy does not change the energy distribution) and bi-Maxwellians (for cases in which it does). These are not the only two possibilities. For instance, there are reasons to be careful of the bi-Maxwellian model for a scenario involving anisotropic heating. If energy is put into a distribution at some $\varepsilon_0$, the process of populating the higher-energy parts of the distribution with $\varepsilon > \varepsilon_0$ tends to rely on collisions between particles with orthogonal velocities (if one imagines equilibration through pairwise collisions, it is necessary for some collisions to leave one of the particles with more energy than either started with). 
These orthogonal collisions also tend to equalize the energies oriented in different directions, so there may be scenarios in which the bi-Maxwellian is a good model for the bulk population of particles but not for the high-energy tails. Interestingly, this also suggests that a distribution occupying a higher-dimensional velocity subspace might be able to populate its high-energy tails more efficiently, since pairs of particles are more likely to have orthogonal velocities. 
In any case, it is important to consider carefully before picking a distribution with which to model the plasma in a particular system, especially given the sensitivity of the fusion yield on the structure of the high-energy tails. 

\begin{acknowledgements}

The authors would like to thank Nicolas Lopez, Ian Ochs, and Xin Zhang for helpful discussions.  This work was supported by NSF PHY-1805316 and NNSA 83228-10966 [Prime No. DOE (NNSA) DE-NA0003764].

\end{acknowledgements}

\section*{Data Availability Statement}

The data that support the findings of this study are available from the corresponding author upon reasonable request.

\bibliographystyle{apsrev4-1} 

\begin{thebibliography}{30}%
	\makeatletter
	\providecommand \@ifxundefined [1]{%
		\@ifx{#1\undefined}
	}%
	\providecommand \@ifnum [1]{%
		\ifnum #1\expandafter \@firstoftwo
		\else \expandafter \@secondoftwo
		\fi
	}%
	\providecommand \@ifx [1]{%
		\ifx #1\expandafter \@firstoftwo
		\else \expandafter \@secondoftwo
		\fi
	}%
	\providecommand \natexlab [1]{#1}%
	\providecommand \enquote  [1]{``#1''}%
	\providecommand \bibnamefont  [1]{#1}%
	\providecommand \bibfnamefont [1]{#1}%
	\providecommand \citenamefont [1]{#1}%
	\providecommand \href@noop [0]{\@secondoftwo}%
	\providecommand \href [0]{\begingroup \@sanitize@url \@href}%
	\providecommand \@href[1]{\@@startlink{#1}\@@href}%
	\providecommand \@@href[1]{\endgroup#1\@@endlink}%
	\providecommand \@sanitize@url [0]{\catcode `\\12\catcode `\$12\catcode
		`\&12\catcode `\#12\catcode `\^12\catcode `\_12\catcode `\%12\relax}%
	\providecommand \@@startlink[1]{}%
	\providecommand \@@endlink[0]{}%
	\providecommand \url  [0]{\begingroup\@sanitize@url \@url }%
	\providecommand \@url [1]{\endgroup\@href {#1}{\urlprefix }}%
	\providecommand \urlprefix  [0]{URL }%
	\providecommand \Eprint [0]{\href }%
	\providecommand \doibase [0]{http://dx.doi.org/}%
	\providecommand \selectlanguage [0]{\@gobble}%
	\providecommand \bibinfo  [0]{\@secondoftwo}%
	\providecommand \bibfield  [0]{\@secondoftwo}%
	\providecommand \translation [1]{[#1]}%
	\providecommand \BibitemOpen [0]{}%
	\providecommand \bibitemStop [0]{}%
	\providecommand \bibitemNoStop [0]{.\EOS\space}%
	\providecommand \EOS [0]{\spacefactor3000\relax}%
	\providecommand \BibitemShut  [1]{\csname bibitem#1\endcsname}%
	\let\auto@bib@innerbib\@empty
	\bibitem [{\citenamefont {Rome}\ and\ \citenamefont {Peng}(1979)}]{Rome1979}%
	\BibitemOpen
	\bibfield  {author} {\bibinfo {author} {\bibfnamefont {J.~A.}\ \bibnamefont
			{Rome}}\ and\ \bibinfo {author} {\bibfnamefont {Y.-K.~M.}\ \bibnamefont
			{Peng}},\ }\href {\doibase 10.1088/0029-5515/19/9/003} {\bibfield  {journal}
		{\bibinfo  {journal} {Nucl. Fusion}\ }\textbf {\bibinfo {volume} {19}},\
		\bibinfo {pages} {1193} (\bibinfo {year} {1979})}\BibitemShut {NoStop}%
	\bibitem [{\citenamefont {Lin-Liu}\ and\ \citenamefont
		{Miller}(1995)}]{LinLiu1995}%
	\BibitemOpen
	\bibfield  {author} {\bibinfo {author} {\bibfnamefont {Y.~R.}\ \bibnamefont
			{Lin-Liu}}\ and\ \bibinfo {author} {\bibfnamefont {R.~L.}\ \bibnamefont
			{Miller}},\ }\href {\doibase 10.1063/1.871315} {\bibfield  {journal}
		{\bibinfo  {journal} {Phys. Plasmas}\ }\textbf {\bibinfo {volume} {2}},\
		\bibinfo {pages} {1666} (\bibinfo {year} {1995})}\BibitemShut {NoStop}%
	\bibitem [{\citenamefont {Cooper}\ \emph {et~al.}(2006)\citenamefont {Cooper},
		\citenamefont {Graves}, \citenamefont {Hirshman}, \citenamefont {Yamaguchi},
		\citenamefont {Narushima}, \citenamefont {Okamura}, \citenamefont
		{Sakakibara}, \citenamefont {Suzuki}, \citenamefont {Watanabe}, \citenamefont
		{Yamada},\ and\ \citenamefont {Yamazaki}}]{Cooper2006}%
	\BibitemOpen
	\bibfield  {author} {\bibinfo {author} {\bibfnamefont {W.~A.}\ \bibnamefont
			{Cooper}}, \bibinfo {author} {\bibfnamefont {J.~P.}\ \bibnamefont {Graves}},
		\bibinfo {author} {\bibfnamefont {S.~P.}\ \bibnamefont {Hirshman}}, \bibinfo
		{author} {\bibfnamefont {T.}~\bibnamefont {Yamaguchi}}, \bibinfo {author}
		{\bibfnamefont {Y.}~\bibnamefont {Narushima}}, \bibinfo {author}
		{\bibfnamefont {S.}~\bibnamefont {Okamura}}, \bibinfo {author} {\bibfnamefont
			{S.}~\bibnamefont {Sakakibara}}, \bibinfo {author} {\bibfnamefont
			{C.}~\bibnamefont {Suzuki}}, \bibinfo {author} {\bibfnamefont {K.~Y.}\
			\bibnamefont {Watanabe}}, \bibinfo {author} {\bibfnamefont {H.}~\bibnamefont
			{Yamada}}, \ and\ \bibinfo {author} {\bibfnamefont {K.}~\bibnamefont
			{Yamazaki}},\ }\href {\doibase 10.1088/0029-5515/46/7/001} {\bibfield
		{journal} {\bibinfo  {journal} {Nucl. Fusion}\ }\textbf {\bibinfo {volume}
			{46}},\ \bibinfo {pages} {683} (\bibinfo {year} {2006})}\BibitemShut
	{NoStop}%
	\bibitem [{\citenamefont {Cooper}\ \emph {et~al.}(2009)\citenamefont {Cooper},
		\citenamefont {Hirshman}, \citenamefont {Merkel}, \citenamefont {Graves},
		\citenamefont {Kisslinger}, \citenamefont {Wobig}, \citenamefont {Narushima},
		\citenamefont {Okamura},\ and\ \citenamefont {Watanabe}}]{Cooper2009}%
	\BibitemOpen
	\bibfield  {author} {\bibinfo {author} {\bibfnamefont {W.~A.}\ \bibnamefont
			{Cooper}}, \bibinfo {author} {\bibfnamefont {S.~P.}\ \bibnamefont
			{Hirshman}}, \bibinfo {author} {\bibfnamefont {P.}~\bibnamefont {Merkel}},
		\bibinfo {author} {\bibfnamefont {J.~P.}\ \bibnamefont {Graves}}, \bibinfo
		{author} {\bibfnamefont {J.}~\bibnamefont {Kisslinger}}, \bibinfo {author}
		{\bibfnamefont {H.~F.~G.}\ \bibnamefont {Wobig}}, \bibinfo {author}
		{\bibfnamefont {Y.}~\bibnamefont {Narushima}}, \bibinfo {author}
		{\bibfnamefont {S.}~\bibnamefont {Okamura}}, \ and\ \bibinfo {author}
		{\bibfnamefont {K.~Y.}\ \bibnamefont {Watanabe}},\ }\href {\doibase
		10.1016/j.cpc.2009.04.006} {\bibfield  {journal} {\bibinfo  {journal}
			{Comput. Phys. Commun.}\ }\textbf {\bibinfo {volume} {180}},\ \bibinfo
		{pages} {1524} (\bibinfo {year} {2009})}\BibitemShut {NoStop}%
	\bibitem [{\citenamefont {Ochs}\ and\ \citenamefont
		{Fisch}(2018)}]{Ochs2018ii}%
	\BibitemOpen
	\bibfield  {author} {\bibinfo {author} {\bibfnamefont {I.~E.}\ \bibnamefont
			{Ochs}}\ and\ \bibinfo {author} {\bibfnamefont {N.~J.}\ \bibnamefont
			{Fisch}},\ }\href {\doibase 10.1063/1.5055568} {\bibfield  {journal}
		{\bibinfo  {journal} {Phys. Plasmas}\ }\textbf {\bibinfo {volume} {25}},\
		\bibinfo {pages} {122306} (\bibinfo {year} {2018})}\BibitemShut {NoStop}%
	\bibitem [{\citenamefont {Choe}\ \emph {et~al.}(1995)\citenamefont {Choe},
		\citenamefont {Chang},\ and\ \citenamefont {Ono}}]{Choe1995}%
	\BibitemOpen
	\bibfield  {author} {\bibinfo {author} {\bibfnamefont {W.}~\bibnamefont
			{Choe}}, \bibinfo {author} {\bibfnamefont {C.~S.}\ \bibnamefont {Chang}}, \
		and\ \bibinfo {author} {\bibfnamefont {M.}~\bibnamefont {Ono}},\ }\href
	{\doibase 10.1063/1.871456} {\bibfield  {journal} {\bibinfo  {journal} {Phys.
				Plasmas}\ }\textbf {\bibinfo {volume} {2}},\ \bibinfo {pages} {2044}
		(\bibinfo {year} {1995})}\BibitemShut {NoStop}%
	\bibitem [{\citenamefont {Maximov}\ \emph {et~al.}(2004)\citenamefont
		{Maximov}, \citenamefont {Anikeev}, \citenamefont {Bagryansky}, \citenamefont
		{Ivanov}, \citenamefont {Lizunov}, \citenamefont {Murakhtin}, \citenamefont
		{Noack},\ and\ \citenamefont {Prikhodko}}]{Maximov2004}%
	\BibitemOpen
	\bibfield  {author} {\bibinfo {author} {\bibfnamefont {V.~V.}\ \bibnamefont
			{Maximov}}, \bibinfo {author} {\bibfnamefont {A.~V.}\ \bibnamefont
			{Anikeev}}, \bibinfo {author} {\bibfnamefont {P.~A.}\ \bibnamefont
			{Bagryansky}}, \bibinfo {author} {\bibfnamefont {A.~A.}\ \bibnamefont
			{Ivanov}}, \bibinfo {author} {\bibfnamefont {A.~A.}\ \bibnamefont {Lizunov}},
		\bibinfo {author} {\bibfnamefont {S.~V.}\ \bibnamefont {Murakhtin}}, \bibinfo
		{author} {\bibfnamefont {K.}~\bibnamefont {Noack}}, \ and\ \bibinfo {author}
		{\bibfnamefont {V.~V.}\ \bibnamefont {Prikhodko}},\ }\href {\doibase
		10.1088/0029-5515/44/4/008} {\bibfield  {journal} {\bibinfo  {journal} {Nucl.
				Fusion}\ }\textbf {\bibinfo {volume} {44}},\ \bibinfo {pages} {542} (\bibinfo
		{year} {2004})}\BibitemShut {NoStop}%
	\bibitem [{\citenamefont {Yamaguchi}\ \emph {et~al.}(2005)\citenamefont
		{Yamaguchi}, \citenamefont {Watanabe}, \citenamefont {Sakakibara},
		\citenamefont {Narushima}, \citenamefont {Narihara}, \citenamefont
		{Tokuzawa}, \citenamefont {Tanaka}, \citenamefont {Yamada}, \citenamefont
		{Osakabe}, \citenamefont {Yamada}, \citenamefont {Kawahata}, \citenamefont
		{Yamazaki},\ and\ \citenamefont {\relax{LHD Experimental
				Group}}}]{Yamaguchi2005}%
	\BibitemOpen
	\bibfield  {author} {\bibinfo {author} {\bibfnamefont {T.}~\bibnamefont
			{Yamaguchi}}, \bibinfo {author} {\bibfnamefont {K.~Y.}\ \bibnamefont
			{Watanabe}}, \bibinfo {author} {\bibfnamefont {S.}~\bibnamefont
			{Sakakibara}}, \bibinfo {author} {\bibfnamefont {Y.}~\bibnamefont
			{Narushima}}, \bibinfo {author} {\bibfnamefont {K.}~\bibnamefont {Narihara}},
		\bibinfo {author} {\bibfnamefont {T.}~\bibnamefont {Tokuzawa}}, \bibinfo
		{author} {\bibfnamefont {K.}~\bibnamefont {Tanaka}}, \bibinfo {author}
		{\bibfnamefont {I.}~\bibnamefont {Yamada}}, \bibinfo {author} {\bibfnamefont
			{M.}~\bibnamefont {Osakabe}}, \bibinfo {author} {\bibfnamefont
			{H.}~\bibnamefont {Yamada}}, \bibinfo {author} {\bibfnamefont
			{K.}~\bibnamefont {Kawahata}}, \bibinfo {author} {\bibfnamefont
			{K.}~\bibnamefont {Yamazaki}}, \ and\ \bibinfo {author} {\bibnamefont
			{\relax{LHD Experimental Group}}},\ }\href {\doibase
		10.1088/0029-5515/45/11/L01} {\bibfield  {journal} {\bibinfo  {journal}
			{Nucl. Fusion}\ }\textbf {\bibinfo {volume} {45}},\ \bibinfo {pages} {L33}
		(\bibinfo {year} {2005})}\BibitemShut {NoStop}%
	\bibitem [{\citenamefont {Qu}\ \emph {et~al.}(2014)\citenamefont {Qu},
		\citenamefont {Fitzgerald},\ and\ \citenamefont {Hole}}]{Qu2014}%
	\BibitemOpen
	\bibfield  {author} {\bibinfo {author} {\bibfnamefont {Z.~S.}\ \bibnamefont
			{Qu}}, \bibinfo {author} {\bibfnamefont {M.}~\bibnamefont {Fitzgerald}}, \
		and\ \bibinfo {author} {\bibfnamefont {M.~J.}\ \bibnamefont {Hole}},\ }\href
	{\doibase 10.1088/0741-3335/56/7/075007} {\bibfield  {journal} {\bibinfo
			{journal} {Plasma Phys. Control. Fusion}\ }\textbf {\bibinfo {volume} {56}},\
		\bibinfo {pages} {075007} (\bibinfo {year} {2014})}\BibitemShut {NoStop}%
	\bibitem [{\citenamefont {Mikkelsen}(1989)}]{Mikkelsen1989}%
	\BibitemOpen
	\bibfield  {author} {\bibinfo {author} {\bibfnamefont {D.~R.}\ \bibnamefont
			{Mikkelsen}},\ }\href {\doibase 10.1088/0029-5515/29/7/003} {\bibfield
		{journal} {\bibinfo  {journal} {Nucl. Fusion}\ }\textbf {\bibinfo {volume}
			{29}},\ \bibinfo {pages} {1113} (\bibinfo {year} {1989})}\BibitemShut
	{NoStop}%
	\bibitem [{\citenamefont {Towner}\ \emph {et~al.}(1992)\citenamefont {Towner},
		\citenamefont {Goldston}, \citenamefont {Hammett}, \citenamefont {Murphy},
		\citenamefont {Phillips}, \citenamefont {Scott}, \citenamefont {Zarnstorff},\
		and\ \citenamefont {Smithe}}]{Towner1992}%
	\BibitemOpen
	\bibfield  {author} {\bibinfo {author} {\bibfnamefont {H.~H.}\ \bibnamefont
			{Towner}}, \bibinfo {author} {\bibfnamefont {R.~J.}\ \bibnamefont
			{Goldston}}, \bibinfo {author} {\bibfnamefont {G.~W.}\ \bibnamefont
			{Hammett}}, \bibinfo {author} {\bibfnamefont {J.~A.}\ \bibnamefont {Murphy}},
		\bibinfo {author} {\bibfnamefont {C.~K.}\ \bibnamefont {Phillips}}, \bibinfo
		{author} {\bibfnamefont {S.~D.}\ \bibnamefont {Scott}}, \bibinfo {author}
		{\bibfnamefont {M.~C.}\ \bibnamefont {Zarnstorff}}, \ and\ \bibinfo {author}
		{\bibfnamefont {D.}~\bibnamefont {Smithe}},\ }\href {\doibase
		10.1063/1.1143630} {\bibfield  {journal} {\bibinfo  {journal} {Rev. Sci.
				Instrum.}\ }\textbf {\bibinfo {volume} {63}},\ \bibinfo {pages} {4753}
		(\bibinfo {year} {1992})}\BibitemShut {NoStop}%
	\bibitem [{\citenamefont {Hay}\ and\ \citenamefont
		{Fisch}(2015)}]{Hay2015ignition}%
	\BibitemOpen
	\bibfield  {author} {\bibinfo {author} {\bibfnamefont {M.~J.}\ \bibnamefont
			{Hay}}\ and\ \bibinfo {author} {\bibfnamefont {N.~J.}\ \bibnamefont
			{Fisch}},\ }\href {\doibase 10.1063/1.4936346} {\bibfield  {journal}
		{\bibinfo  {journal} {Phys. Plasmas}\ }\textbf {\bibinfo {volume} {22}},\
		\bibinfo {pages} {112116} (\bibinfo {year} {2015})}\BibitemShut {NoStop}%
	\bibitem [{\citenamefont {Kiwamoto}\ \emph {et~al.}(1996)\citenamefont
		{Kiwamoto}, \citenamefont {Tatematsu}, \citenamefont {Saito}, \citenamefont
		{Abe}, \citenamefont {Ichimura}, \citenamefont {Inutake}, \citenamefont
		{Yamaguchi}, \citenamefont {Tamano}, \citenamefont {Nakashima}, \citenamefont
		{Shoji}, \citenamefont {Cho}, \citenamefont {Hirata}, \citenamefont {Hojo},
		\citenamefont {Ikeda}, \citenamefont {Ishii}, \citenamefont {Itakura},
		\citenamefont {Katanuma}, \citenamefont {Mase}, \citenamefont {Nagayama},\
		and\ \citenamefont {Yatsu}}]{Kiwamoto1996}%
	\BibitemOpen
	\bibfield  {author} {\bibinfo {author} {\bibfnamefont {Y.}~\bibnamefont
			{Kiwamoto}}, \bibinfo {author} {\bibfnamefont {T.}~\bibnamefont {Tatematsu}},
		\bibinfo {author} {\bibfnamefont {T.}~\bibnamefont {Saito}}, \bibinfo
		{author} {\bibfnamefont {H.}~\bibnamefont {Abe}}, \bibinfo {author}
		{\bibfnamefont {M.}~\bibnamefont {Ichimura}}, \bibinfo {author}
		{\bibfnamefont {M.}~\bibnamefont {Inutake}}, \bibinfo {author} {\bibfnamefont
			{N.}~\bibnamefont {Yamaguchi}}, \bibinfo {author} {\bibfnamefont
			{T.}~\bibnamefont {Tamano}}, \bibinfo {author} {\bibfnamefont
			{Y.}~\bibnamefont {Nakashima}}, \bibinfo {author} {\bibfnamefont
			{M.}~\bibnamefont {Shoji}}, \bibinfo {author} {\bibfnamefont
			{T.}~\bibnamefont {Cho}}, \bibinfo {author} {\bibfnamefont {M.}~\bibnamefont
			{Hirata}}, \bibinfo {author} {\bibfnamefont {H.}~\bibnamefont {Hojo}},
		\bibinfo {author} {\bibfnamefont {K.}~\bibnamefont {Ikeda}}, \bibinfo
		{author} {\bibfnamefont {K.}~\bibnamefont {Ishii}}, \bibinfo {author}
		{\bibfnamefont {A.}~\bibnamefont {Itakura}}, \bibinfo {author} {\bibfnamefont
			{I.}~\bibnamefont {Katanuma}}, \bibinfo {author} {\bibfnamefont
			{A.}~\bibnamefont {Mase}}, \bibinfo {author} {\bibfnamefont {Y.}~\bibnamefont
			{Nagayama}}, \ and\ \bibinfo {author} {\bibfnamefont {K.}~\bibnamefont
			{Yatsu}},\ }\href {\doibase 10.1063/1.871885} {\bibfield  {journal} {\bibinfo
			{journal} {Phys. Plasmas}\ }\textbf {\bibinfo {volume} {3}},\ \bibinfo
		{pages} {578} (\bibinfo {year} {1996})}\BibitemShut {NoStop}%
	\bibitem [{\citenamefont {Kalra}\ \emph {et~al.}(1988)\citenamefont {Kalra},
		\citenamefont {Agrawal},\ and\ \citenamefont {Pandimani}}]{Kalra1988}%
	\BibitemOpen
	\bibfield  {author} {\bibinfo {author} {\bibfnamefont {M.~S.}\ \bibnamefont
			{Kalra}}, \bibinfo {author} {\bibfnamefont {S.}~\bibnamefont {Agrawal}}, \
		and\ \bibinfo {author} {\bibfnamefont {S.}~\bibnamefont {Pandimani}},\
	}\href@noop {} {\bibfield  {journal} {\bibinfo  {journal} {Trans. Amer. Nucl.
				Soc.}\ }\textbf {\bibinfo {volume} {56}},\ \bibinfo {pages} {126} (\bibinfo
		{year} {1988})}\BibitemShut {NoStop}%
	\bibitem [{\citenamefont {Nath}\ \emph {et~al.}(2013)\citenamefont {Nath},
		\citenamefont {Majumdar},\ and\ \citenamefont {Kalra}}]{Nath2013}%
	\BibitemOpen
	\bibfield  {author} {\bibinfo {author} {\bibfnamefont {D.}~\bibnamefont
			{Nath}}, \bibinfo {author} {\bibfnamefont {R.}~\bibnamefont {Majumdar}}, \
		and\ \bibinfo {author} {\bibfnamefont {M.~S.}\ \bibnamefont {Kalra}},\ }\href
	{\doibase 10.1007/s10894-013-9594-0} {\bibfield  {journal} {\bibinfo
			{journal} {J. Fusion Energy}\ }\textbf {\bibinfo {volume} {32}},\ \bibinfo
		{pages} {457} (\bibinfo {year} {2013})}\BibitemShut {NoStop}%
	\bibitem [{\citenamefont {Bosch}\ and\ \citenamefont {Hale}(1992)}]{Bosch1992}%
	\BibitemOpen
	\bibfield  {author} {\bibinfo {author} {\bibfnamefont {H.-S.}\ \bibnamefont
			{Bosch}}\ and\ \bibinfo {author} {\bibfnamefont {G.~M.}\ \bibnamefont
			{Hale}},\ }\href {\doibase 10.1088/0029-5515/32/4/I07} {\bibfield  {journal}
		{\bibinfo  {journal} {Nucl. Fusion}\ }\textbf {\bibinfo {volume} {32}},\
		\bibinfo {pages} {611} (\bibinfo {year} {1992})}\BibitemShut {NoStop}%
	\bibitem [{\citenamefont {Lepage}(1978)}]{Lepage1978}%
	\BibitemOpen
	\bibfield  {author} {\bibinfo {author} {\bibfnamefont {G.~P.}\ \bibnamefont
			{Lepage}},\ }\href {\doibase 10.1016/0021-9991(78)90004-9} {\bibfield
		{journal} {\bibinfo  {journal} {J. Comput. Phys.}\ }\textbf {\bibinfo
			{volume} {27}},\ \bibinfo {pages} {192} (\bibinfo {year} {1978})}\BibitemShut
	{NoStop}%
	\bibitem [{\citenamefont {Nevins}\ and\ \citenamefont
		{Swain}(2000)}]{Nevins2000}%
	\BibitemOpen
	\bibfield  {author} {\bibinfo {author} {\bibfnamefont {W.~M.}\ \bibnamefont
			{Nevins}}\ and\ \bibinfo {author} {\bibfnamefont {R.}~\bibnamefont {Swain}},\
	}\href {\doibase 10.1088/0029-5515/40/4/310} {\bibfield  {journal} {\bibinfo
			{journal} {Nucl. Fusion}\ }\textbf {\bibinfo {volume} {40}},\ \bibinfo
		{pages} {865} (\bibinfo {year} {2000})}\BibitemShut {NoStop}%
	\bibitem [{\citenamefont {Lalousis}\ \emph {et~al.}(2012)\citenamefont
		{Lalousis}, \citenamefont {F\"oldes},\ and\ \citenamefont
		{Hora}}]{Lalousis2012}%
	\BibitemOpen
	\bibfield  {author} {\bibinfo {author} {\bibfnamefont {P.}~\bibnamefont
			{Lalousis}}, \bibinfo {author} {\bibfnamefont {I.~B.}\ \bibnamefont
			{F\"oldes}}, \ and\ \bibinfo {author} {\bibfnamefont {H.}~\bibnamefont
			{Hora}},\ }\href {\doibase 10.1017/S0263034611000875} {\bibfield  {journal}
		{\bibinfo  {journal} {Laser and Particle Beams}\ }\textbf {\bibinfo {volume}
			{30}},\ \bibinfo {pages} {233} (\bibinfo {year} {2012})}\BibitemShut
	{NoStop}%
	\bibitem [{\citenamefont {Hora}\ \emph {et~al.}(2015)\citenamefont {Hora},
		\citenamefont {Korn}, \citenamefont {Giuffrida}, \citenamefont {Margarone},
		\citenamefont {Picciotto}, \citenamefont {Krasa}, \citenamefont {Jungwirth},
		\citenamefont {Ullschmied}, \citenamefont {Lalousis}, \citenamefont
		{Eliezer}, \citenamefont {Miley}, \citenamefont {Moustaizis},\ and\
		\citenamefont {Mourou}}]{Hora2015}%
	\BibitemOpen
	\bibfield  {author} {\bibinfo {author} {\bibfnamefont {H.}~\bibnamefont
			{Hora}}, \bibinfo {author} {\bibfnamefont {G.}~\bibnamefont {Korn}}, \bibinfo
		{author} {\bibfnamefont {L.}~\bibnamefont {Giuffrida}}, \bibinfo {author}
		{\bibfnamefont {D.}~\bibnamefont {Margarone}}, \bibinfo {author}
		{\bibfnamefont {A.}~\bibnamefont {Picciotto}}, \bibinfo {author}
		{\bibfnamefont {J.}~\bibnamefont {Krasa}}, \bibinfo {author} {\bibfnamefont
			{K.}~\bibnamefont {Jungwirth}}, \bibinfo {author} {\bibfnamefont
			{J.}~\bibnamefont {Ullschmied}}, \bibinfo {author} {\bibfnamefont
			{P.}~\bibnamefont {Lalousis}}, \bibinfo {author} {\bibfnamefont
			{S.}~\bibnamefont {Eliezer}}, \bibinfo {author} {\bibfnamefont {G.~H.}\
			\bibnamefont {Miley}}, \bibinfo {author} {\bibfnamefont {S.}~\bibnamefont
			{Moustaizis}}, \ and\ \bibinfo {author} {\bibfnamefont {G.}~\bibnamefont
			{Mourou}},\ }\href {\doibase 10.1017/S0263034615000634} {\bibfield  {journal}
		{\bibinfo  {journal} {Laser and Particle Beams}\ }\textbf {\bibinfo {volume}
			{33}},\ \bibinfo {pages} {607} (\bibinfo {year} {2015})}\BibitemShut
	{NoStop}%
	\bibitem [{\citenamefont {Eliezer}\ \emph {et~al.}(2016)\citenamefont
		{Eliezer}, \citenamefont {Hora}, \citenamefont {Korn}, \citenamefont
		{Nissim},\ and\ \citenamefont {Val}}]{Eliezer2016}%
	\BibitemOpen
	\bibfield  {author} {\bibinfo {author} {\bibfnamefont {S.}~\bibnamefont
			{Eliezer}}, \bibinfo {author} {\bibfnamefont {H.}~\bibnamefont {Hora}},
		\bibinfo {author} {\bibfnamefont {G.}~\bibnamefont {Korn}}, \bibinfo {author}
		{\bibfnamefont {N.}~\bibnamefont {Nissim}}, \ and\ \bibinfo {author}
		{\bibfnamefont {J.~M.~M.}\ \bibnamefont {Val}},\ }\href {\doibase
		10.1063/1.4950824} {\bibfield  {journal} {\bibinfo  {journal} {Phys.
				Plasmas}\ }\textbf {\bibinfo {volume} {23}},\ \bibinfo {pages} {050704}
		(\bibinfo {year} {2016})}\BibitemShut {NoStop}%
	\bibitem [{\citenamefont {Putvinski}\ \emph {et~al.}(2019)\citenamefont
		{Putvinski}, \citenamefont {Ryutov},\ and\ \citenamefont
		{Yushmanov}}]{Putvinski2019}%
	\BibitemOpen
	\bibfield  {author} {\bibinfo {author} {\bibfnamefont {S.~V.}\ \bibnamefont
			{Putvinski}}, \bibinfo {author} {\bibfnamefont {D.~D.}\ \bibnamefont
			{Ryutov}}, \ and\ \bibinfo {author} {\bibfnamefont {P.~N.}\ \bibnamefont
			{Yushmanov}},\ }\href {\doibase 10.1088/1741-4326/ab1a60} {\bibfield
		{journal} {\bibinfo  {journal} {Nucl. Fusion}\ }\textbf {\bibinfo {volume}
			{59}},\ \bibinfo {pages} {076018} (\bibinfo {year} {2019})}\BibitemShut
	{NoStop}%
	\bibitem [{\citenamefont {Hora}\ \emph {et~al.}(2020)\citenamefont {Hora},
		\citenamefont {Miley}, \citenamefont {Eliezer},\ and\ \citenamefont
		{Nissim}}]{Hora2020}%
	\BibitemOpen
	\bibfield  {author} {\bibinfo {author} {\bibfnamefont {H.}~\bibnamefont
			{Hora}}, \bibinfo {author} {\bibfnamefont {G.~H.}\ \bibnamefont {Miley}},
		\bibinfo {author} {\bibfnamefont {S.}~\bibnamefont {Eliezer}}, \ and\
		\bibinfo {author} {\bibfnamefont {N.}~\bibnamefont {Nissim}},\ }\href
	{\doibase 10.1016/j.hedp.2019.100739} {\bibfield  {journal} {\bibinfo
			{journal} {High Density Phys.}\ }\textbf {\bibinfo {volume} {35}},\ \bibinfo
		{pages} {100739} (\bibinfo {year} {2020})}\BibitemShut {NoStop}%
	\bibitem [{\citenamefont {Newcomb}(1981)}]{Newcomb1981}%
	\BibitemOpen
	\bibfield  {author} {\bibinfo {author} {\bibfnamefont {W.~A.}\ \bibnamefont
			{Newcomb}},\ }\href {\doibase 10.1017/S0022377800010904} {\bibfield
		{journal} {\bibinfo  {journal} {J. Plasma Phys.}\ }\textbf {\bibinfo {volume}
			{26}},\ \bibinfo {pages} {529} (\bibinfo {year} {1981})}\BibitemShut
	{NoStop}%
	\bibitem [{\citenamefont {Lehnert}(1971)}]{Lehnert1971}%
	\BibitemOpen
	\bibfield  {author} {\bibinfo {author} {\bibfnamefont {B.}~\bibnamefont
			{Lehnert}},\ }\href {\doibase 10.1088/0029-5515/11/5/010} {\bibfield
		{journal} {\bibinfo  {journal} {Nucl. Fusion}\ }\textbf {\bibinfo {volume}
			{11}},\ \bibinfo {pages} {485} (\bibinfo {year} {1971})}\BibitemShut
	{NoStop}%
	\bibitem [{\citenamefont {Pastukhov}(1974)}]{Pastukhov1974}%
	\BibitemOpen
	\bibfield  {author} {\bibinfo {author} {\bibfnamefont {V.~P.}\ \bibnamefont
			{Pastukhov}},\ }\href {\doibase 10.1088/0029-5515/14/1/001} {\bibfield
		{journal} {\bibinfo  {journal} {Nucl. Fusion}\ }\textbf {\bibinfo {volume}
			{14}},\ \bibinfo {pages} {3} (\bibinfo {year} {1974})}\BibitemShut {NoStop}%
	\bibitem [{\citenamefont {Chernin}\ and\ \citenamefont
		{Rosenbluth}(1978)}]{Chernin1978}%
	\BibitemOpen
	\bibfield  {author} {\bibinfo {author} {\bibfnamefont {D.~P.}\ \bibnamefont
			{Chernin}}\ and\ \bibinfo {author} {\bibfnamefont {M.~N.}\ \bibnamefont
			{Rosenbluth}},\ }\href {\doibase 10.1088/0029-5515/18/1/008} {\bibfield
		{journal} {\bibinfo  {journal} {Nucl. Fusion}\ }\textbf {\bibinfo {volume}
			{18}},\ \bibinfo {pages} {47} (\bibinfo {year} {1978})}\BibitemShut {NoStop}%
	\bibitem [{\citenamefont {Cohen}\ \emph {et~al.}(1978)\citenamefont {Cohen},
		\citenamefont {Rensink}, \citenamefont {Cutler},\ and\ \citenamefont
		{Mirin}}]{Cohen1978}%
	\BibitemOpen
	\bibfield  {author} {\bibinfo {author} {\bibfnamefont {R.~H.}\ \bibnamefont
			{Cohen}}, \bibinfo {author} {\bibfnamefont {M.~E.}\ \bibnamefont {Rensink}},
		\bibinfo {author} {\bibfnamefont {T.~A.}\ \bibnamefont {Cutler}}, \ and\
		\bibinfo {author} {\bibfnamefont {A.~A.}\ \bibnamefont {Mirin}},\ }\href
	{\doibase 10.1088/0029-5515/18/9/005} {\bibfield  {journal} {\bibinfo
			{journal} {Nucl. Fusion}\ }\textbf {\bibinfo {volume} {18}},\ \bibinfo
		{pages} {1229} (\bibinfo {year} {1978})}\BibitemShut {NoStop}%
	\bibitem [{\citenamefont {Bekhtenev}\ \emph {et~al.}(1980)\citenamefont
		{Bekhtenev}, \citenamefont {Volosov}, \citenamefont {Pal'chikov},
		\citenamefont {Pekker},\ and\ \citenamefont {\relax{Yu.}
			N.~Yudin}}]{Bekhtenev1980}%
	\BibitemOpen
	\bibfield  {author} {\bibinfo {author} {\bibfnamefont {A.~A.}\ \bibnamefont
			{Bekhtenev}}, \bibinfo {author} {\bibfnamefont {V.~I.}\ \bibnamefont
			{Volosov}}, \bibinfo {author} {\bibfnamefont {V.~E.}\ \bibnamefont
			{Pal'chikov}}, \bibinfo {author} {\bibfnamefont {M.~S.}\ \bibnamefont
			{Pekker}}, \ and\ \bibinfo {author} {\bibnamefont {\relax{Yu.} N.~Yudin}},\
	}\href {\doibase 10.1088/0029-5515/20/5/007} {\bibfield  {journal} {\bibinfo
			{journal} {Nucl. Fusion}\ }\textbf {\bibinfo {volume} {20}},\ \bibinfo
		{pages} {579} (\bibinfo {year} {1980})}\BibitemShut {NoStop}%
	\bibitem [{\citenamefont {Zhang}\ \emph {et~al.}(2019)\citenamefont {Zhang},
		\citenamefont {Elliott}, \citenamefont {Maan}, \citenamefont {Boyle},
		\citenamefont {Kaita},\ and\ \citenamefont {Majeski}}]{Zhang2019}%
	\BibitemOpen
	\bibfield  {author} {\bibinfo {author} {\bibfnamefont {X.}~\bibnamefont
			{Zhang}}, \bibinfo {author} {\bibfnamefont {D.~B.}\ \bibnamefont {Elliott}},
		\bibinfo {author} {\bibfnamefont {A.}~\bibnamefont {Maan}}, \bibinfo {author}
		{\bibfnamefont {D.~P.}\ \bibnamefont {Boyle}}, \bibinfo {author}
		{\bibfnamefont {R.}~\bibnamefont {Kaita}}, \ and\ \bibinfo {author}
		{\bibfnamefont {R.}~\bibnamefont {Majeski}},\ }\href {\doibase
		10.1016/j.nme.2019.02.027} {\bibfield  {journal} {\bibinfo  {journal} {Nucl.
				Mater. Energy}\ }\textbf {\bibinfo {volume} {19}},\ \bibinfo {pages} {250}
		(\bibinfo {year} {2019})}\BibitemShut {NoStop}%
\end{thebibliography}
\providecommand{\noopsort}[1]{}\providecommand{\singleletter}[1]{#1}%

\end{document}